# A perturbation density functional theory for the competition between inter and intramolecular association

Bennett D. Chapman[a,1], Alejandro J. García-Cuéllar[b] and Walter G. Chapman[a]

[a]Department of Chemical and Biomolecular Engineering, Rice University, 6100 S. Main, Houston, Texas 77005

[b]Department of Mechanical Engineering, Tecnológico de Monterrey, Av. Eugenio Garza Sada 2501, Monterrey, N.L. 64849, México

**Abstract**

Using the framework of Wertheim's thermodynamic perturbation theory we develop the first density functional theory which accounts for intramolecular association in chain molecules. To test the theory new Monte Carlo simulations are performed at a fluid solid interface for a 4 segment chain which can both intra and intermolecularly associate. The theory and simulation results are found to be in excellent agreement. It is shown that the inclusion of intramolecular association can have profound effects on interfacial properties such as interfacial tension and the partition coefficient.

---

[1] Author to whom correspondence should be addressed
Email: bennettd1980@gmail.com

**Introduction:**

Hydrogen bonding (association) plays an integral role in our everyday lives.[1] From the remarkable properties of water to the folding of proteins[2] hydrogen bonding is key to our very existence. Modeling associating fluids is complicated by highly directional asymmetric interactions; for this reason the development of accurate statistical mechanical based theories for associating fluids lagged behind that of simple fluids with spherically symmetric potentials. In the 1980's Wertheim[3-7] developed a theory capable of accurately describing associating fluids by introducing the highly directional interactions at an early point in the theory. By introducing a multi-density formalism, where each bonding state of a molecule is treated as a distinct species, Wertheim was able to rewrite the statistical mechanics of associating fluids in a form which was very amiable to approximation. One such approximation, Wertheim's thermodynamic perturbation theory[4, 6, 7] (TPT), has proven remarkably successful. In TPT the change in free energy due to association is obtained as a perturbation to a hard sphere reference fluid. TPT is typically used as a first order perturbation theory (TPT1), and provides a basis for the SAFT[8, 9] equation of state; SAFT has found widespread use in both industry and academia.[10]

One key approximation introduced in TPT is the neglect of all graphs with rings of association bonds. For most systems this approximation will introduce a small or nonexistent error, however, molecules such as glycol ethers[11] show a significant degree of intramolecular association which affects the thermodynamics of the system. To account for the possibility of intramolecular association Sear and Jackson[12] modified TPT by adding a ring graph to the fundamental graph sum. In a separate approach Ghonasgi and Chapman[13, 14] developed a theory to account for intramolecular association; their theory was found to be in excellent agreement with molecular simulations.

In addition to homogeneous systems, TPT, has proven versatile and accurate in the description of inhomogeneous systems.[15] By letting the association energy become infinitely large, complex polyatomic molecules can be constructed allowing the development of polymer density functional theories[16-21], DFT's, in the framework of TPT. In addition, if some association energies are allowed to remain finite, DFT's capable of describing associating polyatomic molecules can be developed.[22-24] In these associating DFT's the possibility of intramolecular association has been neglected. If we are to develop an accurate DFT for the description of interfacial systems involving glycol ethers, or to accurately describe protein folding[2], the possibility of intramolecular association must be accounted for.

In this work we will develop a DFT capable of describing molecules which can both intra and intermolecularly associate. We will follow Sear and Jackson[12] and introduce a ring graph in the fundamental graph sum to account for intramolecular association. With this free energy functional, we will construct and minimize a grand potential which will allow us to obtain the inhomogeneous density profiles. As a test of the theory, we perform new Monte Carlo simulations for a 4-mer chain which can both intra and intermolecularly associate near a hard wall. The theory is shown to be in excellent agreement with simulation. We show that interfacial properties such as interfacial tension and the partition coefficient are strongly affected by intramolecular association.

## II: Theory

In this section we will introduce the type of molecules we want to study, the potential model and develop the Helmholtz free energy and segment densities. Here we will consider linear fully flexible molecules of length $m$ consisting of hard spheres (segments) where each location on the chain is occupied by a certain species of segment. Each segment has two association sites A (red) and B (green) as shown in Fig. 1. The interaction potential between segments $\beta$ and $\gamma$ is given as the sum of a hard sphere and association potential

$$\phi^{(\beta,\gamma)}(12) = \phi_{HS}^{(\beta,\gamma)}(r_{12}) + \phi_{AB}^{(\beta,\gamma)}(12) + \phi_{BA}^{(\beta,\gamma)}(12) \tag{1}$$

The notation (1) represents the position $\vec{r}_1$ and orientation $\Omega_1$ of a spherical segment, $r_{12}$ is the distance between the segments and $\phi_{HS}^{(\beta,\gamma)}(r_{12})$ is the hard sphere potential

$$\phi_{HS}^{(\beta,\gamma)}(r_{12}) = \begin{cases} 0 & r_{12} \geq \sigma^{(\beta,\gamma)} \\ \infty & r_{12} < \sigma^{(\beta,\gamma)} \end{cases} \tag{2}$$

where $\sigma^{(\beta,\gamma)}$ is the cross species diameter. The association potential $\phi_{AB}^{(\beta,\gamma)}(12)$ is that of a conical site[9]

$$\phi_{AB}^{(i,j)}(i,j) = \begin{cases} -\varepsilon_{AB}^{(ij)}, & r_{ij} < r_c; \theta_{Ai} < \theta_c; \theta_{Bj} < \theta_c \\ 0 & otherwise \end{cases} \tag{3}$$

Where $\theta_{Ai}$ is the angle between the vector from the center of segment $i$ to site A and the vector $\vec{r}_{ij}$, $\theta_c$ is the cutoff angle beyond which association is not allowed and $r_c$ is the cutoff radius which is the maximum separation between two segments where association can occur. There is no association between sites of the same type, that is $\varepsilon_{AA}^{(ij)} = \varepsilon_{BB}^{(ij)} = 0$. To create the chain, the limit of complete association is taken, $\varepsilon_{AB}^{(ij)} \to \infty$, for all association bonds internal to the chain while leaving the association energy between the A association site on segment 1 and the B association site on segment $m$ finite and adjustable. As illustrated in Fig.2, both intermolecular and intramolecular association is allowed.

In Wertheim's theory each bonding state of a molecule is treated as a distinct species. The density of species $\beta$ bonded at a set of sites $\alpha$ at location 1 in the fluid is $\rho_\alpha^{(\beta)}(1)$. For the 2 site fluid the total density of component $\beta$ will be the sum of the segments which are bonded at both sites A and B, those bonded at sites A or B and those which are not bonded

$$\rho^{(\beta)}(1) = \rho_{AB}^{(\beta)}(1) + \rho_A^{(\beta)}(1) + \rho_B^{(\beta)}(1) + \rho_o^{(\beta)}(1) \qquad (4)$$

where $\rho_o^{(\beta)}(1)$ is the monomer density. We will also use a set of density parameters

$$\sigma_A^{(\beta)}(1) = \rho_A^{(\beta)}(1) + \rho_o^{(\beta)}(1)$$

$$\sigma_B^{(\beta)}(1) = \rho_B^{(\beta)}(1) + \rho_o^{(\beta)}(1) \qquad (5)$$

In TPT the change in free energy due to association is,

$$\beta A^{Wertheim} = \sum_{\beta=1}^m \int \left( \rho^{(\beta)}(1) \ln \frac{\rho_o^{(\beta)}(1)}{\rho^{(\beta)}(1)} + \rho^{(\beta)}(1) + Q^{(\beta)}(1) \right) d(1) - \Delta c^{(o)} \qquad (6)$$

Equation (6) is written for molecules with fixed bond angles between association sites. To allow for distributions of bond angles, bond angles $\alpha$ are treated as internal variables and bond angle distribution functions $\xi(\alpha)$ can be introduced.[7] In the fully flexible limit non adjacent segments on the chain can overlap and $\xi(\alpha) = 1/2$ at which point Eq. (6) is recovered. Introducing the bond angle correlation functions will not change the form of the results, so for notational simplicity we will not use this formality. The form of the final equations is valid for fully and semi – flexible chains, one simply needs to enforce any bond angle constraints.

For a two site fluid $Q^{(\beta)}(1)$ is given as

$$Q^{(\beta)}(1) = -\sigma_A^{(\beta)}(1) - \sigma_B^{(\beta)}(1) + \frac{\sigma_A^{(\beta)}(1)\,\sigma_B^{(\beta)}(1)}{\rho_o^{(\beta)}(1)} \tag{7}$$

The fundamental graph sum $\Delta c^{(o)}$ for this type of molecule can be written as the sum of contributions from chain formation, ring formation and intermolecular association (polymerization)

$$\Delta c^{(o)} = \Delta c_{chain}^{(o)} + \Delta c_{ring}^{(o)} + \Delta c_{poly}^{(o)} \tag{8}$$

Where $\Delta c_{chain}^{(o)}$ and $\Delta c_{poly}^{(o)}$ are given by[7]

$$\Delta c_{chain}^{(o)} = \sum_{\beta=1}^{m-1} \int \sigma_A^{(\beta)}(1)\Delta^{(\beta,\beta+1)}(12)\sigma_B^{(\beta+1)}(2)d(1)d(2) \tag{9}$$

$$\Delta c_{poly}^{(o)} = \int \sigma_B^{(1)}(1)\Delta^{poly}(12)\sigma_A^{(m)}(2)d(1)d(2) \tag{10}$$

The term $\Delta^{(i,j)}(12) = y^{(i,j)}(12)F^{(i,j)}(12)$ where $y^{(i,j)}(12)$ is the inhomogeneous cavity correlation function of the reference system and $F^{(i,j)}(12) = \exp\left(-\beta\phi_{HS}^{(i,j)}(r_{12})\right)f_{AB}^{(i,j)}(12)$. The contribution due to intramolecular association is given by Sear and Jackson's ring graph[12]

$$\Delta c_{ring}^{(o)} = \int \tilde{\Delta}^{Ring}(1...m)\prod_{\in=1}^{m}\rho_o^{(\in)}(\in)d(\in) \tag{11}$$

Where

$$\tilde{\Delta}^{Ring}(1...m) = \tilde{\Delta}^{Chain}(1...m)\Delta^{ring}(1,m) \tag{12}$$

and,

$$\tilde{\Delta}^{Chain}(1...m) = \Delta^{(12)}(12)\Delta^{(23)}(23)...\Delta^{(m-1,m)}(m-1,m) \tag{13}$$

We minimize the free energy with respect to monomer densities $\rho_o^{(j)}(j)$ to obtain

$$\frac{\rho^{(j)}(j)}{\rho_o^{(j)}(j)} = \frac{\sigma_A^{(j)}(j)\sigma_B^{(j)}(j)}{\left(\rho_o^{(j)}(j)\right)^2} + \int \tilde{\Delta}^{Ring}(1...m)\prod_{\in\neq j}^{m}\rho_o^{(\in)}(\in)d(\in) \tag{14}$$

Now minimizing with respect to the $\sigma_A^{(j)}(j)$ and $\sigma_B^{(j)}(j)$ for chain forming association sites

$$\frac{\sigma_B^{(j)}(j)}{\rho_o^{(j)}(j)} - 1 = \int \Delta^{(j,j+1)}(j,j+1)\sigma_B^{(j+1)}(j+1)d(j+1) \tag{15}$$

$$\frac{\sigma_A^{(j)}(j)}{\rho_o^{(j)}(j)} - 1 = \int \sigma_A^{(j-1)}(j-1)\Delta^{(j-1,j)}(j-1,j)d(j-1) \tag{16}$$

For site A on segment 1 and site B on segment $m$

$$\frac{\sigma_A^{(1)}(1)}{\rho_o^{(1)}(1)} - 1 = \int \sigma_A^{(m)}(2)\Delta^{poly}(12)d(2) \tag{17}$$

$$\frac{\sigma_B^{(m)}(1)}{\rho_o^{(m)}(1)} - 1 = \int \sigma_B^{(1)}(1)\Delta^{poly}(12)d(2) \tag{18}$$

Using Eqns. (6) through (18) the Helmholtz free energy can be written as

$$\beta A^{Wertheim} = \sum_{\beta=1}^{m} \int \left( \rho^{(\beta)}(1)\ln\frac{\rho_o^{(\beta)}(1)}{\rho^{(\beta)}(1)} + \rho^{(\beta)}(1) - \sigma_B^{(\beta)}(1) \right) d(1)$$
$$- \int \left( \rho^{(1)}(1) - \frac{\sigma_A^{(1)}(1)\,\sigma_B^{(1)}(1)}{\rho_o^{(1)}(1)} \right) d(1) \tag{19}$$

Equation (19) was originally derived by Sear and Jackson[12] in the development of a bulk equation of state, however Eq. (19) is general for inhomogeneous systems. Writing Eq. (14) for segment 1

$$\frac{\rho^{(1)}(1)}{\rho_o^{(1)}(1)} = \frac{\sigma_A^{(1)}(1)\,\sigma_B^{(1)}(1)}{\left(\rho_o^{(1)}(1)\right)^2} + I_{ring}^{(1)}(1) \tag{20}$$

Where

$$I_{ring}^{(j)}(j) = \int \widetilde{\Delta}^{Ring}(1...m) \prod_{\in \neq j}^{m} \rho_o^{(\in)}(\in) d(\in)$$

(21)

Combining Eqns. (17) and (20)

$$\sigma_B^{(1)}(1) = \frac{\rho^{(1)}(1) - \rho_o^{(1)}(1) I_{ring}^{(1)}(1)}{1 + \int \sigma_A^{(m)}(2) \Delta^{poly}(12) d(2)}$$

(22)

Dividing each side of Eq. (22) by $\rho^{(1)}(1)$ we obtain

$$X_A(1) = X_A^{poly}(1) \left(1 - \chi_{ring}(1)\right)$$

(23)

Where $X_A(1)$ is the fraction of component 1 *not* bonded at site A, $X_A(1) = \sigma_B^{(1)}(1) / \rho^{(1)}(1)$, similarly for site B on segment $m$ $X_B(1) = \sigma_A^{(m)}(1) / \rho^{(m)}(1)$. The ring fraction $\chi_{ring}(1)$ is the fraction of species 1 which is bonded intramolecularly to species $m$

$$\chi_{ring}(1) = \frac{\rho_o^{(1)}(1)}{\rho^{(1)}(1)} I_{ring}^{(1)}(1)$$

(24)

and $X_A^{poly}(1)$ is the fraction of component 1 not bonded at site A if only intermolecular association were possible

$$X_A^{poly}(1) = \frac{1}{1 + \int \rho^{(m)}(2) X_B(2) \Delta^{poly}(12) d(2)}$$

(25)

and the equivalent fraction for site B on segment $m$

$$X_B^{poly}(1) = \frac{1}{1 + \int \rho^{(1)}(2) X_A(2) \Delta^{poly}(12) d(2)} \tag{26}$$

Now we wish to take the limit of complete association of all chain forming sites. In the limit of complete association of chain forming sites monomer densities become small, so the second term on the left hand side of Eqns. (15) and (16) can be neglected. Equations (15) and (16) are now used to recursively eliminate the density parameters $\sigma_A^{(i)}$ and $\sigma_B^{(i)}$ associated with chain forming sites in Eq. (14). The resulting segmental densities are

$$\frac{\rho^{(j)}(j)}{\rho_o^{(j)}(j)} = \int \frac{1}{X_A^{poly}(1)} \widetilde{\Delta}^{chain}(1...m) \frac{1}{X_B^{poly}(m)} \prod_{\in \neq j}^{m} \rho_o^{(\in)}(\in) d(\in)$$
$$+ \int \widetilde{\Delta}^{Ring}(1...m) \prod_{\in \neq j}^{m} \rho_o^{(\in)}(\in) d(\in) \tag{27}$$

We will find it necessary to employ a two point chain density $\rho^{(j,k)}(j,k)$; to obtain this quantity we first note that the densities in Eq. (27) can be obtained through functional derivatives of a generating functional $\Delta \widetilde{c}^{(o)}$

$$\frac{\rho^{(j)}(j)}{\rho_o^{(j)}(j)} = \frac{\delta \Delta \widetilde{c}^{(o)}}{\delta \rho_o^{(j)}(j)} \tag{28}$$

The functional $\Delta \widetilde{c}^{(o)}$ is given by

$$\Delta \widetilde{c}^{(o)} = \int \frac{1}{X_A^{poly}(1)} \widetilde{\Delta}^{chain}(1...m) \frac{1}{X_B^{poly}(m)} \prod_{\in = 1}^{m} \rho_o^{(\in)}(\in) d(\in)$$
$$+ \int \widetilde{\Delta}^{Ring}(1...m) \prod_{\in = 1}^{m} \rho_o^{(\in)}(\in) d(\in) \tag{29}$$

Using the generating functional $\Delta \tilde{c}^{(o)}$ we can also obtain the two point chain density as

$$\frac{\rho^{(j,k)}(j,k)}{\rho_o^{(j)}(j)\rho_o^{(k)}(k)} = \frac{\delta^2 \Delta \tilde{c}^{(o)}}{\delta \rho_o^{(k)}(k)\delta \rho_o^{(j)}(j)} \tag{30}$$

Evaluating Eq. (30)

$$\frac{\rho^{(j,k)}(j,k)}{\rho_o^{(j)}(j)\rho_o^{(k)}(k)} = \int \frac{1}{X_A^{poly}(1)} \tilde{\Delta}^{chain}(1...m) \frac{1}{X_B^{poly}(m)} \prod_{\substack{\in \neq j \\ \in \neq k}}^{m} \rho_o^{(\in)}(\in) d(\in)$$

$$+ \int \tilde{\Delta}^{Ring}(1...m) \prod_{\substack{\in \neq j \\ \in \neq k}}^{m} \rho_o^{(\in)}(\in) d(\in) \tag{31}$$

Comparing Eqns. (27) and (31) it is clear that

$$\rho^{(j)}(j) = \int \rho^{(j,k)}(j,k) d(k) \tag{32}$$

To complete the theory we must obtain equations for the $m$ unknown $\rho_o^{(j)}(j)$'s. To determine these densities density functional theory will be employed. In density functional theory we define a grand potential at fixed chemical potential $\mu$, volume $V$ and temperature $T$ subject to an external field $V_{ext}^{(\beta)}(\vec{r}')$

$$\Omega[\{\rho^{(\beta)}(\vec{r})\}] = A[\{\rho^{(\beta)}(\vec{r})\}] - \sum_{\beta=1}^{m} \int d\vec{r}' \rho^{(\beta)}(\vec{r}') \left(\mu^{(\beta)} - V_{ext}^{(\beta)}(\vec{r}')\right) \tag{33}$$

Where

$$\rho^{(\gamma)}(\vec{r}_1) = \int \rho^{(\gamma)}(1)d\Omega_1 \qquad (34)$$

is the density integrated over all orientations; there is a similar relation for $\rho_o^{(\gamma)}(\vec{r}_1)$.

Minimization of the grand potential with respect to the segment densities yields the set of Euler – Lagrange equations.

$$\frac{\delta A[\{\rho^{(\beta)}(\vec{r})\}]}{\delta\rho^{(\gamma)}(\vec{r})} = \mu^{(\gamma)} - V_{ext}^{(\gamma)} \qquad \forall \ \gamma = 1, m \qquad (35)$$

The solution of this set of equations will yield the needed monomer densities. The Helmholtz free energy is given as

$$A[\{\rho^{(\beta)}\}] = A^{id}[\{\rho^{(\beta)}\}] + A^{Wertheim}[\{\rho^{(\beta)}\}] + A^{HS}[\{\rho^{(\beta)}\}] \qquad (36)$$

where $A^{id}[\{\rho^{(\beta)}\}]$, $A^{Wertheim}[\{\rho^{(\beta)}\}]$ and $A^{HS}[\{\rho^{(\beta)}\}]$ are the contributions from ideal gas and excess contributions due to chain formation / association (Eq.19) and hard sphere repulsions. The functional derivatives of the ideal gas term is known exactly

$$\frac{\delta\beta A^{ideal}[\{\rho^{(\beta)}\}]}{\delta\rho^{(\gamma)}(\vec{r})} = \ln \rho^{(\gamma)}(\vec{r}) \qquad (37)$$

For the hard sphere term we use Rosenfeld's fundamental measure theory[25]

$$\frac{\delta\beta A^{HS}[\{\rho^{(\beta)}\}]}{\delta\rho^{(\gamma)}(\vec{r})} = \int d\vec{r}_1 \frac{\delta\Phi^{ex,hs}[\{n^{(j)}\}]}{\delta\rho^{(\gamma)}(\vec{r})} \qquad (38)$$

To obtain the contribution due to $A^{Wertheim}[\{\rho^{(\beta)}\}]$ we take the functional derivative of Eq. (6) and enforce the limit of complete association of chain forming sites, we obtain

$$\frac{\delta \beta A^{Wertheim}}{\delta \rho^{(j)}(\vec{r}_j)} = \ln \frac{\rho_o^{(j)}(\vec{r}_j)}{\rho^{(j)}(\vec{r}_j)} - \frac{\delta \Delta c^{(o)}}{\delta \rho^{(j)}(\vec{r}_j)} \tag{39}$$

where

$$\frac{\delta \Delta c^{(o)}}{\delta \rho^{(j)}(\vec{r}_j)} = \sum_{\beta=1}^{m-1} \int \rho^{(\beta,\beta+1)}(\vec{r}_1,\vec{r}_2) \frac{\delta \ln y^{(\beta,\beta+1)}(\vec{r}_1,\vec{r}_2)}{\delta \rho^{(j)}(\vec{r}_j)} d\vec{r}_1 d\vec{r}_2$$

$$+ \int \rho^{(1)}(\vec{r}_1)\rho^{(m)}(\vec{r}_2) X_A(\vec{r}_1) X_B(\vec{r}_2) \Delta^{poly}(\vec{r}_1,\vec{r}_2) \frac{\delta \ln \Delta^{poly}(\vec{r}_1,\vec{r}_2)}{\delta \rho^{(j)}(\vec{r}_j)} d\vec{r}_1 d\vec{r}_2$$

$$+ \int \rho_{ring}^{(1,m)}(\vec{r}_1,\vec{r}_m) \frac{\delta \ln \Delta^{ring}(\vec{r}_1,\vec{r}_m)}{\delta \rho^{(j)}(\vec{r}_j)} d\vec{r}_1 d\vec{r}_m \tag{40}$$

The intermolecular association strength $\Delta^{poly}(\vec{r}_1,\vec{r}_2)$ is given as[26]

$$\Delta^{poly}(\vec{r}_1,\vec{r}_2) = \kappa_{AB} f_{AB}^{(1,m)}(\vec{r}_1,\vec{r}_2) g^{(1m)}(\vec{r}_1,\vec{r}_2) \tag{41}$$

The two point density $\rho_{ring}^{(1,m)}(\vec{r}_1,\vec{r}_m)$ is given by

$$\frac{\rho_{ring}^{(1,m)}(\vec{r}_1,\vec{r}_m)}{\rho_o^{(1)}(\vec{r}_1)\rho_o^{(m)}(\vec{r}_m)} = \int \tilde{\Delta}^{Ring}(1...m) \prod_{\in=2}^{m-1} \rho_o^{(\in)}(\vec{r}_\in) d\vec{r}_\in \tag{42}$$

We can relate the two point ring density to the total segment 1 density and ring fraction

$$\rho^{(1)}(\vec{r}_1)\chi_{ring}(\vec{r}_1) = \int \rho_{ring}^{(1,m)}(\vec{r}_1,\vec{r}_m) d\vec{r}_m \tag{43}$$

Using Eqns. (35) – (39) we can solve for the monomer densities

$$\rho_o^{(j)}(\vec{r}_j) = \exp\left[\lambda^{(j)}(\vec{r}_j) + \beta\mu^{(j)}\right] \qquad (44)$$

where

$$\lambda^{(j)}(\vec{r}_j) = \frac{\delta\Delta c^{(o)}}{\delta\rho^{(j)}(\vec{r}_j)} - \frac{\delta\beta A^{HS}}{\delta\rho^{(j)}(\vec{r}_j)} - \beta V_{ext}^{(j)}(\vec{r}_j) \qquad (45)$$

Using Eq. (44) to eliminate the monomer densities in Eq. (27) we obtain

$$\rho^{(j)}(\vec{r}_j) = \exp[\lambda^{(j)}(\vec{r}_j) + \beta\mu_M]\left(I_{chain}^{(j)}(\vec{r}_j) + I_{Ring}^{(j)}(\vec{r}_j)\right) \qquad (46)$$

Where $\mu_M = \sum_{k=1}^{m}\mu^{(k)}$ is the molecular chemical potential, $I_{Ring}^{(j)}(\vec{r}_j)$ is given by

$$I_{Ring}^{(j)}(\vec{r}_j) = \int \widetilde{\Delta}^{Ring}(1...m)\prod_{\in\neq j}^{m}\exp[\lambda^{(\in)}(\vec{r}_\in)]d\vec{r}_\in \qquad (47)$$

For fully flexible chains the chain integral $I_{chain}^{(j)}(\vec{r}_j)$ is factored and evaluated with recursion relations

$$I_{chain}^{(j)}(\vec{r}_j) = I_1^{(j)}(\vec{r}_j)I_2^{(j)}(\vec{r}_j) \qquad (48)$$

The $I_1^{(j)}(\vec{r}_j)'s$ are evaluated using the following recursion relations

$$I_1^{(1)}(\vec{r}_1) = \frac{1}{X_A^{poly}(\vec{r}_1)} \tag{49}$$

and for $j > 1$,

$$I_1^{(j)}(\vec{r}_j) = \int \Delta^{(j-1,j)}(\vec{r}_{j-1}, \vec{r}_j) \exp\left[\lambda^{(j-1)}(\vec{r}_{j-1})\right] I_1^{(j-1)}(\vec{r}_{j-1}) d\vec{r}_{j-1} \tag{50}$$

Here $\Delta^{(i,j)}(\vec{r}_i, \vec{r}_j) = \Lambda y^{(i,j)}(\vec{r}_i, \vec{r}_j) \dfrac{\delta\left(\left|\vec{r}_i - \vec{r}_j\right| - \sigma^{(i,j)}\right)}{4\pi\left(\sigma^{(i,j)}\right)^2}$ where $\Lambda$ is the infinitely large magnitude of

the chain forming Mayer functions, as will be seen this term cancels exactly with an identical

term in the chemical potential. Similarly for the $I_2^{(j)}(\vec{r}_j)'s$

$$I_2^{(m)}(\vec{r}_m) = \frac{1}{X_B^{poly}(\vec{r}_m)} \tag{51}$$

and for $j < m$,

$$I_2^{(j)}(\vec{r}_j) = \int \Delta^{(j,j+1)}(\vec{r}_j, \vec{r}_{j+1}) \exp\left[\lambda^{(j+1)}(\vec{r}_{j+1})\right] I_2^{(j+1)}(\vec{r}_{j+1}) d\vec{r}_{j+1} \tag{52}$$

All that remains is the approximation of the reference system correlation functions. Kierlik and

Rosinberg[27] approximated the inhomogeneous hard sphere pair correlation function as a first order functional Taylor Series in density around the homogeneous result. If we took this path we would first be required to solve for the two point density Eq. (31) and then integrate through Eq.

(32) to obtain the segment densities. A simpler approach[20, 28] which has proven to yield accurate results[26, 29, 30] is to approximate the reference pair cavity correlation function as the average of the potential of mean force

$$\ln y^{(\alpha,\beta)}(\vec{r}_1,\vec{r}_2) = \frac{1}{2}\ln\left\{y^{(\alpha,\beta)}(\vec{r}_1)\,y^{(\alpha,\beta)}(\vec{r}_2)\right\} \tag{53}$$

Where the $y^{(\alpha,\beta)}(\vec{r}_1)$ are evaluated by using the bulk result at an average density

$$\overline{\rho}^{(\gamma)}(\vec{r}_1) = \frac{3}{4\pi\sigma^3}\int_{|r_1-r_2|<\sigma} d\vec{r}_2\,\rho^{(\gamma)}(\vec{r}_2) \tag{54}$$

Using Eqns. (23), (25), (32), (43), (53) and assuming a symmetric molecule we can rewrite Eq. (40) as

$$\frac{\delta\Delta c^{(o)}}{\delta\rho^{(j)}(\vec{r}_j)} = \frac{1}{2}\sum_{\beta=1}^{m-1}\int\left(\rho^{(\beta)}(\vec{r}_1)+\rho^{(\beta+1)}(\vec{r}_1)\right)\frac{\delta\ln y^{(\beta,\beta+1)}(\vec{r}_1)}{\delta\rho^{(j)}(\vec{r}_j)}\,d\vec{r}_1$$
$$+\int\rho^{(1)}(\vec{r}_1)\left(1-X_A(\vec{r}_1)\right)\frac{\delta\ln y^{(1,m)}(\vec{r}_1)}{\delta\rho^{(j)}(\vec{r}_j)}\,d\vec{r}_1 \tag{55}$$

Equation (55) completes the density functional theory for the competition between inter and intramolecular association in associating chain fluids. Unfortunately, the ring integral in Eq. (47) is irreducible and cannot be factorized. For large flexible rings direct numerical evaluation of this integral by quadrature will be computationally impractical. One possible resolution would be to evaluate the ring integral by single chain Monte Carlo simulation.[31, 32] This will be the subject of a future study and will not be discussed further here. The appendix gives a detailed discussion of methods to evaluate the ring integral.

For the systems studied in the paper the Calculational method is as follows. A bulk density $\rho$ is specified and the bulk $X_A$ is calculated by[13]

$$\left(\frac{1}{X_A}\right)^3 + \left(\rho\Delta^{poly} - \Delta^{ring} - 1\right)\left(\frac{1}{X_A}\right)^2 - 2\rho\Delta^{poly}\left(\frac{1}{X_A}\right) - \left(\rho\Delta^{poly}\right)^2 = 0 \tag{56}$$

Using this $X_A$ the bulk $X_A^{poly}$ can be calculated using Eq. (25) which allows for the calculation of the bulk $\chi_{ring}$ through Eq. (23). We can solve for the excess contribution to the chemical potential due to chain formation and association $\mu_{Werheim}$; the result for a homonuclear molecule is

$$\beta\mu_{Werheim} = \ln X_A + \ln X_A^{poly} - (m-1)\ln y - (m-1)\rho_m\frac{\partial\ln y}{\partial\rho_m}$$
$$- (1-X_A)\rho_m\frac{\partial\ln y}{\partial\rho_m} - (m-1)\ln\rho - (m-1)\ln\Lambda \tag{57}$$

where $\rho_m = m\rho$ is the total segment density. The last term in Eq. (57) containing the association strength for chain forming bonds, $\Lambda$, gives an infinite contribution, however this term cancels exactly with the $\Lambda$'s contained in the chain forming Mayor functions. Now Eqns. (46) for the density profiles and an additional equation for the ring fraction

$$\chi_{ring}(\vec{r}) = \frac{I_{ring}^{(1)}(\vec{r})}{I_{ring}^{(1)}(\vec{r}) + I_{chain}^{(1)}(\vec{r})} \tag{58}$$

are solved using a Picard iteration where the initial guess for the density and ring fraction profiles are the bulk values at each point in the domain.

### III: Simulation

To test the theory, DFT calculations and molecular simulations will be compared for the classical case of a fluid in a planar slit pore of width $H$ with walls located in the $xy$ plane subject to the external potential

$$V_{ext}^{(j)}\left(z_j\right) = \begin{cases} \infty & if \quad z_j < 0 \; or \; z_j > H \\ 0 & otherwise \end{cases} \tag{59}$$

We will use the molecular model of Ghonasgi and Chapman[13] who considered 4 tangentially bonded hard sphere segments with association sites located on the first and fourth segments. The association sites are arranged such that the vector from the center of the associating segment to the association site is always at a $90°$ angle to the vector which points from the center of the associating segment to the center of the neighboring segment on the chain, see Fig. 3. The chain molecules interact with the potential given by Eq. (1) with the cutoff parameters chosen as $r_c = 1.1\sigma$ and $\theta_c = 27^o$.

Molecular simulations are performed in the NVT ensemble using the general method described in ref [13]. A total of 287 chain molecules were simulated in a box with two hard walls on opposite sides. For the other four sides, periodic boundary conditions were applied. Maximum displacement and angle change parameters are adjusted in each simulation run to allow for an overall 30-40% rate of acceptance. The simulations where carried out for $2(10)^6$ cycles, where a cycle consists of an attempt to displace and reorient all molecules once. The results for the density profiles and bonding fractions were obtained after the molecular configurations were sufficiently equilibrated. The system was said to be equilibrated once the fractions of component 1 bonded intramolecularly $\chi_{ring}(z)$ and intermolecularly $\chi_{inter}(z)$ had achieved steady values throughout the pore. At the high association energies $\varepsilon/kT = 7$ and 8 and at a packing fraction of $\eta = 0.3$, $\chi_{ring}(z)$ and $\chi_{inter}(z)$ did not stabilize sufficiently throughout the entire pore over the length of our simulations.

**IV: Results**

In this section we compare density functional theory (DFT) calculations to the Monte Carlo (MC) simulations discussed in section III for the case of an associating 4 – mer chain near a hard wall. When the 4-mer chain self associates into a ring, the short range repulsions of the molecule will keep the associated ring in a nearly planar configuration. Hence, to a good approximation, we can approximate the ring integral as that of a planar ring; see the appendix for approximation of $I_{ring}^{(j)}$ for this case. All calculations performed in this section are for the case $\varepsilon^{inter} = \varepsilon^{intra} = \varepsilon$ and all density profiles are scaled by the average density of a segment in the pore $\rho_{ave}$. There are two segment types in this molecule; the end segments with association sites will be called type 1, and the middle segments will be called type 2. Figure 4 compares MC and DFT density profile calculations at an average packing fraction in the pore of $\eta = 0.1$. At this low density both end and middle segments are depleted from the wall due to a loss of configurational entropy near wall contact, with the wall contact value of the end segment always larger than that of the middle segment. As association energy increases, the density of segment 2 in contact with the wall remains approximately constant while that of segment 1 decreases. The decrease in the wall density of segment 1 is the result of a loss of configurations where this segment can be near the wall when association into rings or longer m – mers occurs. Figure 5 compares MC and DFT density profile calculations for an average pore fraction $\eta = 0.2$. Like the $\eta = 0.1$ case, increasing the association energy results in a decrease in the density wall contact value of segment 1. The theory is very accurate in predicting the density profile of the associating segment, while it is less accurate for the middle segments. In general these types of perturbation density functional theories will be most accurate for end type segments due to the fact that the density profile of an end segment is closer to that of the reference hard sphere fluid than that of a middle segment.[27]

Figure 6 gives density profiles for η = 0.3. At this packing fraction hard sphere packing effects result in an enhancement in density at wall contact. As the association energy is increased to $\varepsilon = 6kT$, the density contact value of segment 1 decreases while that of segment two remains constant.

In addition to density profiles we can also calculate the fraction of segment type 1 bonded (associated) intramoleculary $\chi_{ring}$, and the fraction of segment type 1 bonded intermolecularly $\chi_{inter}$; Fig. 7 compares DFT and MC calculations of these quantities for average system packing fractions of η = 0.1 and 0.2. In general, the fractions bonded intramoleculary show a maximum around z = σ and approximately obtain there bulk value at wall contact while the fractions bonded intermolecularly show a steady decrease as the wall is approached. Intermolecular association is hindered near wall contact due to the fact that there are less ways that two chains can position and orient themselves such that association may occur. The situation for intramolecular association is quite different. The degree of intramolecular association depends on the probability that the two ends of the chain are positioned such that association can occur. At wall contact approximately half of the chain configurations which can lead to intramolecular association in the bulk will be available, however when segment 1 is in contact with the wall only half as many chain configurations in total, as compared to the bulk, will be available; hence the ratio of these quantities at wall contact should approximately yield the bulk result giving a contact value of $\chi_{ring}$ nearly that of the bulk fluid. The MC and DFT predictions are in excellent agreement.

The thermodynamics of the system depends on the fraction of component 1 not bonded $X_A(z)$. Figure 8 compares MC and DFT (solid lines) calculations for $X_A(z)$ at packing fractions of η = 0.1, 0.2 and 0.3. For comparison we have included DFT calculations (dashed lines) where the possibility of intramolecular association was neglected. We see that the current DFT is in excellent agreement with simulation, while DFT's which do not include the possibility of intramolecular association under predict the amount of association in the system. For systems where intramolecular association can occur the current DFT is clearly superior to previous versions of DFT.

With the current DFT we can study how the competition between inter and intramolecular association affects partitioning at a solid / fluid interface. Figure 9 presents partition coefficients at packing fractions of $\eta = 0.1$ and 0.2. At $\eta = 0.1$ when only intermolecular association is considered the partition coefficient continually decreases as association energy is increased (T decreased) due to the fact that the chain molecules are associating into longer m – mers which excludes associated clusters from the wall. However, when intramolecular association is accounted for we see a minimum in the partition coefficient near $\varepsilon / kT = 11$ where the partition coefficient begins to increase with association energy. The minimum in the partition coefficient results from the fact that at low densities and high association energies (low T) intramolecular association dominates[13]; breaking intermolecular association bonds to form intramolecular bonds results in smaller associated clusters which can more easily approach the wall, resulting in an increase in the partition coefficient. At a packing fraction of $\eta = 0.2$ this minimum disappears. Increasing density further to $\eta = 0.3$ does not change the qualitative dependence of the partition coefficient on association energy observed in the $\eta = 0.2$ case.

Interestingly, the results in Figure 9 look very similar to the compressibility factors calculated by MC simulations by Ghonasgi and Chapman[13]; they studied the bulk behavior of this system. The link between the partition coefficient and the bulk compressibility factor is the wall contact theorem which states that the bulk pressure is equal to the wall contact value of the density

$$\frac{P}{kT} = \sum_{i=1}^{4} \rho_i \left( z = 0 \right) \tag{60}$$

Using Eq. (61) we calculated the compressibility factor $Z = \dfrac{P}{\rho kT}$ and compared the results to the simulations of Ghonasgi and Chapman[13] at a bulk packing fraction $\eta_{bulk} = 0.1093$, Fig. 10. We see that the MC and DFT calculations are in good agreement. The minimum in the compressibility factor results from trading intermolecular association bonds for intramolecular association bonds. This results in smaller clusters of associated $4 - $mers and a corresponding increase in the compressibility factor.

Also important in many applications is the interfacial tension $\gamma$ of the solid / fluid interface, where the interfacial tension is calculated as the surface excess grand potential per area of interface $A$

$$\gamma = \frac{\Omega - \Omega_{bulk}}{A} \tag{61}$$

Figure 11 presents DFT calculations for $\gamma$ at packing fractions of $\eta = 0.1$ and 0.2. At $\eta = 0.2$, increasing association energy (decreasing T) results in an increase in $\gamma$ at all energies considered. This increase in $\gamma$ results from attractions between the molecules becoming more significant, so more energy is required to separate the molecules to form the interface; the lower $\gamma$ obtained when intramolecular association is accounted for stems from the fact that molecules which are associated into rings have no attractions to the other molecules in the system. At $\eta = 0.1$ there is still a continues increase in $\gamma$, as association energy is increased, when intramolecular association is neglected, however when intramolecular association is accounted for there is a maximum near $\varepsilon / kT = 8$ and then $\gamma$ begins to decrease. This behavior is analogous to that observed in the partition coefficient, however the maximum in $\gamma$ is located at a lower energy than the minimum $K$, suggesting that the interfacial tension is more affected by ring formation than the partition coefficient. Increasing density further to $\eta = 0.3$ does not change the qualitative dependence of $\gamma$ on association energy observed in the $\eta = 0.2$ case.

**Conclusions:**

We have developed the first density functional theory for chain molecules capable of intramolecular and intermolecular association. As a test, we performed NVT Monte Carlo simulations for a 4 – mer in a slit pore. The theory was shown to be in excellent agreement with simulation results. It was shown that inclusion of intramolecular association can result in drastic qualitative changes to properties such as interfacial tension and the partition coefficient; this behavior cannot be captured with previous versions of DFT.

**<u>Acknowledgements</u>**

The financial support for this work was provided by the Robert A. Welch Foundation (Grant No. C-1241) and by the National Science Foundation (CBET – 0756166).

**Appendix: Calculation of $I_{ring}^{(j)}(z_j)$**

In this appendix methods to evaluate the ring integral $I_{ring}^{(j)}(z_j)$ in planar $1-D$ systems will be discussed. We begin with the intramolecular association strength averaged over segment orientations

$$\Delta^{ring}(\vec{r}_1, \vec{r}_m) = \left\langle f_{AB}^{(intra)}(1m) g^{(1,m)}(1m) \right\rangle_{\Omega_1, \Omega_m} = C \kappa_{AB} y^{(1m)}(\vec{r}_1, \vec{r}_m) F^{(intra)} \xi(r_{1m}) \tag{A1}$$

Where

$$\xi(r_{1m}) = \begin{cases} 1 & \sigma \leq r_{1m} \leq r_c \\ 0 & otherwise \end{cases} \tag{A2}$$

and $C$ is a normalization factor. When evaluating the ring integral $I_{ring}^{(j)}(z_j)$ we are essentially counting the number of configurations the ring can take with segment $j$ at $z_j$ and the ring located in the field created by the other molecules in the fluid and the external potential; for each ring configuration the intramolecular association strength $\Delta^{ring}(\vec{r}_1, \vec{r}_m)$ controls if segments 1 and m located at locations $\vec{r}_1$ and $\vec{r}_m$ in the fluid associate to form a ring. As written, Eq. (47) is for a freely jointed ring where non adjacent segments along the ring can overlap.

For planar systems with inhomogeneities in the $z$ direction the density is a function of $z$ only and we can rewrite the ring integral as

$$I_{Ring}^{(j)}(z_j) = C\kappa_{AB}F^{(intra)}\int D_{ring}(z_1...z_m)\Psi(z_1...z_m)\prod_{\in \neq j}^{m}\exp[\lambda^{(\in)}(z_j)]dz_j \qquad (A3)$$

The function $\Psi(z_1...z_m)$ is a purely geometric quantity given by

$$\Psi(z_1...z_m) = \frac{1}{A}\int \delta(r_{12}-\sigma)...\delta(r_{m-1,m}-\sigma)\xi(r_{1m})\prod_{i=1}^{m}dA_i \qquad (A4)$$

$\Psi(z_1...z_m)$ can be referenced to the location of a segment such that it is independent of the absolute $z$ position in the pore. The integral $\Psi(z_1...z_m)$ is performed once and stored for use. $D_{ring}(z_1...z_m)$ is the product of $m$ cavity correlation functions.

For flexible $4-mer$ chains the homogeneous $\Delta^{ring}$ is known[13]

$$\Delta^{ring} = F^{(intra)}D \qquad (A5)$$

Where

$$D = \frac{2\times10^{-4}}{(1-\eta)^2} \qquad (A6)$$

and $\eta$ is the bulk packing fraction. Normalizing the ring integral to this homogeneous result we find the constant $C$ for a 4 segment chain

$$\frac{1}{C} = \frac{\kappa_{AB}\,yA}{DV}\int dz_1...dz_4\Psi(z_1...z_4) \qquad (A7)$$

Where $V$ is volume and $y$ is the bulk cavity correlation function.

Now as a test we compare DFT calculations to MC simulations of ring fractions for a fluid which can only intramolecularly associate; Fig. 12 shows these results. The theory and simulation are in fair agreement. The theory predicts good ring fraction contact values, however it under predicts the ring peak located near $z = \sigma$ and does not capture the dips near $z = 2\sigma$. These deficiencies arise from the fully flexible treatment of the ring integral. The fully flexible treatment should be sufficient for larger rings, however the self avoiding associated $4 - mer$ ring is sufficiently rigid that the fully flexible treatment of the ring integral will incur error.

One solution is to evaluate both the ring and chain integrals such that no intra – molecule segment overlap is allowed (self – avoiding). An alternative solution which is computationally simpler and faster than the self – avoiding case is to treat the ring integral as rigid with segments 1 and 4 bonded at contact. That is

$$
\begin{aligned}
\Psi_{rigid}\left(z_1 \ldots z_4\right) = \frac{1}{A} \int \delta\left(r_{12} - \sigma\right)\delta\left(r_{23} - \sigma\right)\delta\left(r_{43} - \sigma\right) \\
\times \delta\left(r_{41} - \sigma\right)\delta\left(r_{13} - \sqrt{2}\sigma\right)\delta\left(r_{24} - \sqrt{2}\sigma\right)\prod_{i=1}^{4} dA_i
\end{aligned}
\tag{A8}
$$

Evaluating the ring integral this way will under predict the number of molecular configurations that can lead to ring formation due to the fact that the actual ring has flexibility and association occurs within a shell of thickness $\left(r_c - \sigma\right)$; to correct for this fact we simply include the probability $W$ in $\Delta^{ring}$ for the probability that the chain is in a configuration where the two end segments can associate

$$
\Delta^{ring}\left(\vec{r}_1, \vec{r}_4\right) = C\kappa_{AB}\, y^{(1m)}\left(\vec{r}_1, \vec{r}_4\right)F^{(\text{intra})}W\left(\vec{r}_1\right)
\tag{A9}
$$

Where $W(\vec{r}_1)$ is the probability that in a system with chains of length $m = 4$ and bulk density $\rho$ that if we anchor segment 1 at a position $\vec{r}_1$ that segment 4 will be in a position where intramolecular association can occur; that is $r_{14} \geq \sigma$ and $r_{14} \leq r_c$. We will approximate this quantity as

$$W(\vec{r}_1) = \frac{\int\limits_{r_{14} \geq \sigma, r_{14} \leq r_c} \rho^{(1,4)}_{chain\ ref}(\vec{r}_1, \vec{r}_4) d\vec{r}_4}{\int\limits_{all\ configurations} \rho^{(1,4)}_{chain\ ref}(\vec{r}_1, \vec{r}_4) d\vec{r}_4} = \frac{\int\limits_{r_{14} \geq \sigma, r_{14} \leq r_c} \rho^{(1,4)}_{chain\ ref}(\vec{r}_1, \vec{r}_4) d\vec{r}_4}{\rho^{(1)}_{chain\ ref}(\vec{r}_1)} = \frac{I^{shell}_{chain\ ref}(\vec{r}_1)}{I^{(1)}_{2,chain\ ref}(\vec{r}_1)} \tag{A10}$$

The function $\rho^{(1,4)}_{chain\ ref}(\vec{r}_1, \vec{r}_4)$ is the two point chain density of the non-associating fully flexible chain reference system.

$$\rho^{(1,4)}_{chain\ ref}(\vec{r}_1, \vec{r}_4) = \rho^{(1,4)}(\vec{r}_1, \vec{r}_4)\Big|_{non-associating} \tag{A11}$$

For the $1 - D$ system the integral over the bonding shell of a segment 1 sphere at position $\vec{r}_1$ in the fully flexible reference fluid is

$$I^{shell}_{chain\ ref}(z_1) = \int D_{chain}(z_1 ... z_4) \Psi(z_1 ... z_4) \prod_{\epsilon=2}^{4} \exp[\lambda^{(\epsilon)}_{chain\ ref}(z_\epsilon)] dz_\epsilon \tag{A12}$$

Where $\Psi(z_1 ... z_4)$ is given by Eq. (A4) and $D_{chain}$ is the product of 3 cavity correlation functions. Since $W(z_1)$ is a functional of the chain reference system density profile we can say

$$\frac{\delta \ln \Delta^{ring}(z_1, z_4)}{\delta \rho^{(k)}(z)} = \frac{\delta \ln y^{(1,4)}(z_1, z_4)}{\delta \rho^{(k)}(z)} \tag{A13}$$

We have calculated $W(z)/W_{bulk}$ for a freely jointed 4-mer chain near a hard wall for packing fractions of $\eta = 0.1$, 0.2 and 0.3. These results are presented in Fig. 13. At each density we see a distinct maximum located near $z = \sigma$ and at wall contact the probability is approximately equal to its bulk value. We note that for $\eta = 0.3$ the function $W(z)/W_{bulk}$ has an odd curvature in the region $\sigma / 2 < z < \sigma$.

In molecular simulation two segments can be considered "bonded" even if the association energy is zero. Figure 14 presents simulation results for ring fractions in the non – associating chain reference system at a packing fraction of $\eta = 0.3$. As can be seen the odd shape present in $W(z)$ is also present in this quantity, showing that this is indeed a feature of the chain reference system.

As a test we calculated ring fractions for a 4 –mer chain which can only intramolecularly associate, $\varepsilon^{inter} = 0$, and compared these results to molecular simulation; the results can be seen in Fig. 15. For packing fractions of $\eta = 0.1$ and 0.2 the theoretical results are in excellent agreement with simulation. For $\eta = 0.3$ the theoretical results are in good agreement with the simulation data over most of the domain however the peak in the theoretical calculations near $z \sim \sigma$ has an odd shape. This odd shape is the result of the curvature of the reference system $W(z)$ at this density as discussed above. Overall the agreement with simulation is much better than the fully flexible case.

To obtain improved results at $\eta = 0.3$ we can restrict the chain integral $I_{chain}^{(k)}$ and ring integral $I_{ring}^{(k)}$ such that no intramolecular overlaps are allowed (self avoiding). The chain integral will no longer be able to be factored, Eq. (48) will no longer be valid, and additional multi – dimensional integrals will need to be performed. Since we will not assume that the associated ring is rigid with segments 1 and 4 bonded at contact we will no longer need the reference system probability $W(z)$. The methods to develop $I_{chain}^{(k)}(z)$ and $I_{ring}^{(k)}(z)$ for the self avoiding case is similar to the development of the fully flexible ring integral Eq. (A3) except now additional constraints are added. In the interest of brevity these equations will not be derived here. For $\eta = 0.1$ and 0.2 nearly the same results are obtained for ring fractions as the rigid case. Figure 16 shows the results for ring fractions of an intramoleculary associating fluid an average packing fraction of $\eta = 0.3$. The results are in excellent agreement with simulation. The self avoiding method gives the most accurate results at high density; however, the rigid ring method is computationally faster. For this reason we will employ the rigid ring method to study the competition between intra and intermolecular association.

**Figure Captions:**

**Figure 1:** Formation of associating chain molecules of length m from spherical building blocks.

**Figure 2:** Intermolecular and intramolecular association of associating chain molecules

**Figure 3:** Diagram of associating 4 – mer

**Figure 4:** Density profiles for associating 4 – mer with an average packing fraction η = 0.1. Curves are theoretical calculations (solid – associating end segment 1, dashed – non associating center segment 2) and symbols give Monte Carlo results (red – middle segment, black – end segment)

**Figure 5:** Same as Figure 4 with η = 0.2

**Figure 6:** Same as Figure 4 with η = 0.3

**Figure 7:** Comparison of DFT (curves) and MC (symbols) calculations of the fractions of segment type 1 bonded intermolecularly, $\chi_{inter}$, and the fraction of segment type 1 bonded intramoleculary $\chi_{ring}$.

**Figure 8:** Comparison of current DFT (solid lines), DFT with neglect of intramolecular association (dashed lines) and MC (symbols) calculations for fraction of segment type 1 not bonded $X_A(z)$ for average pore packing fractions of η = 0.1, 0.2 and 0.3

**Figure 9:** Partition coefficient $K = \int\limits_{0}^{4\sigma} dz \frac{\rho(z)/\rho_{bulk}}{4\sigma}$ for packing fractions of $\eta = 0.1$ top and $\eta = 0.2$ bottom. Curves give theoretical predictions (red – current DFT, blue neglecting intramolecular association) and symbols give MC simulations (red squares – both intra and intermolecular association, blue triangles – intermolecular association only)

**Figure 10:** Compressibility factor calculated through the wall contact theorem compared to the MC simulations of Ghonasgi and Chapman[13]. Curves and symbols have same meaning as in Fig. 9

**Figure 11:** Interfacial tension of solid / fluid interface at packing fractions of $\eta = 0.1$ and $0.2$. Curves have same meaning as Fig. 9

**Figure 12:** Fraction of segment 1 bonded intramolecularly $\chi_{ring}$ (curves – DFT , symbols – NVT simulation) for a fluid which is only allowed to intramoleculary associate (no intermolecular association) and the rings are freely jointed

**Figure 13:** Calculation of $W(z)$ for $\eta = 0.1$, $0.2$ and $0.3$

**Figure 14:** Fraction of component 1 not bonded in the non – associating reference system calculated by NVT simulation for a packing fraction of $\eta = 0.3$

**Figure 15:** Same as Fig. 12 except associated rings are assumed rigid and planar

**Figure 16:** Same as bottom panel of Fig. 12 for $\eta = 0.3$ except chains and associated rings are self avoiding

**Figure 1:**

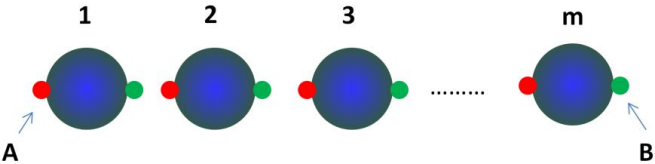

**Figure 2:**

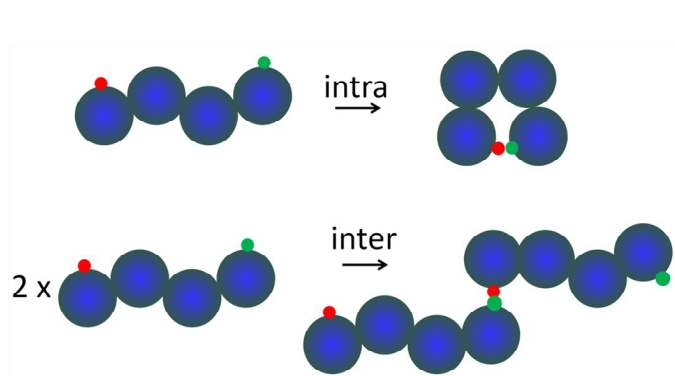

**Figure 3:**

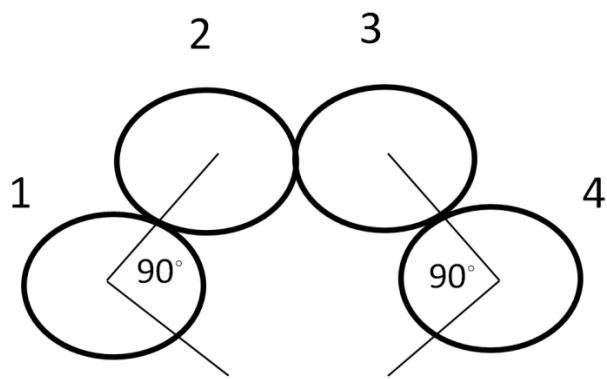

**Figure 4:**

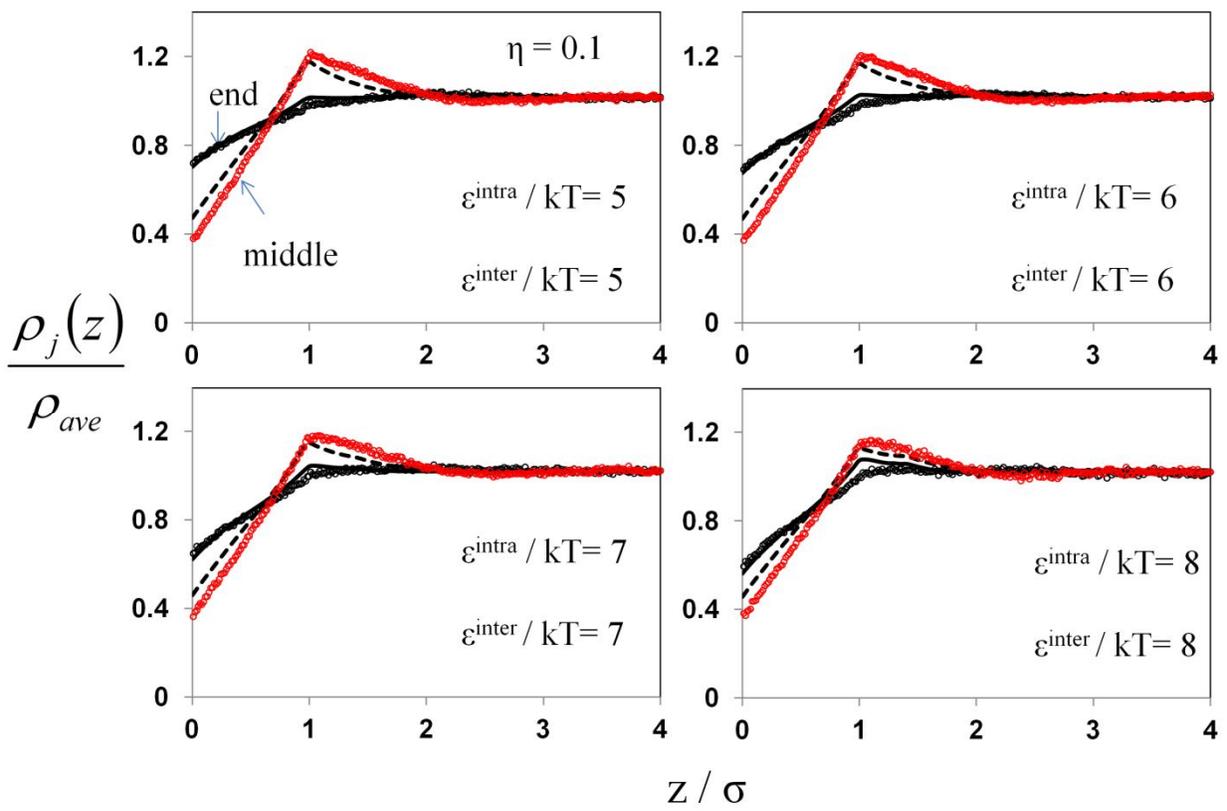

**Figure 5:**

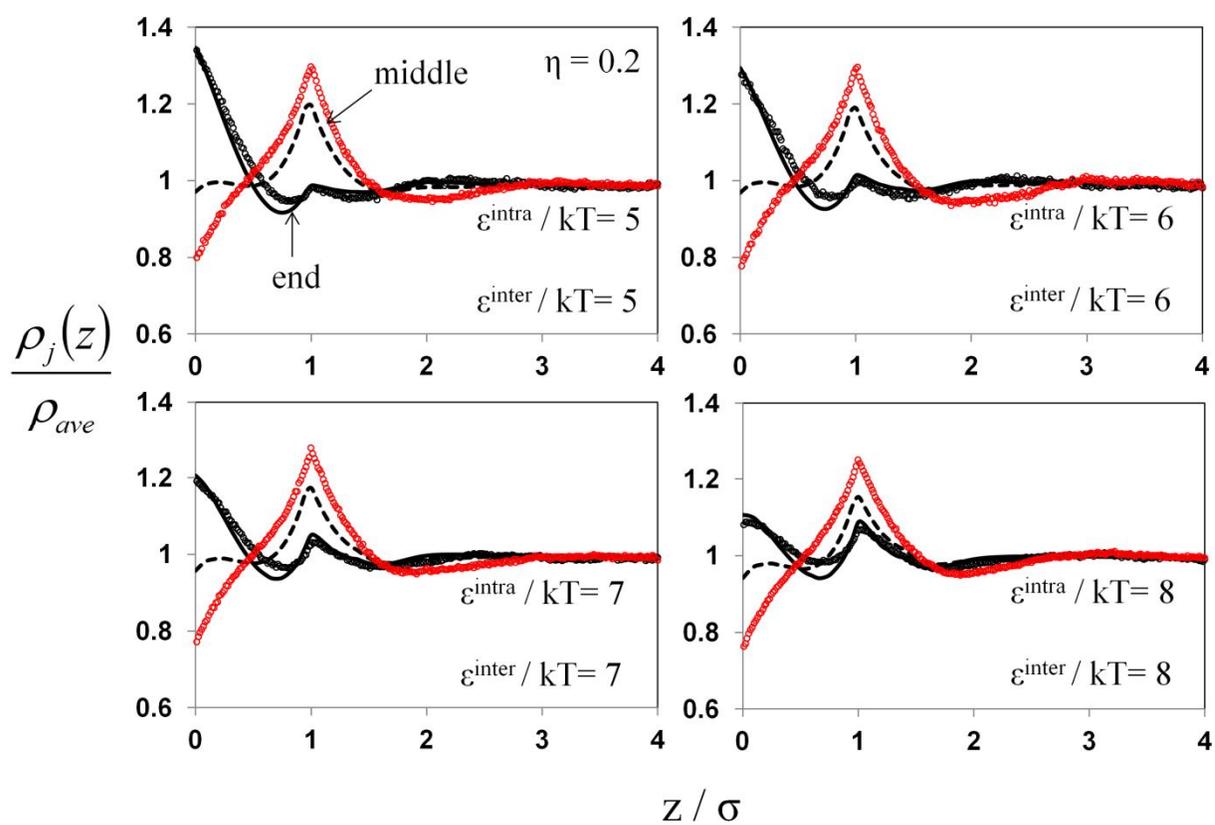

**Figure 6:**

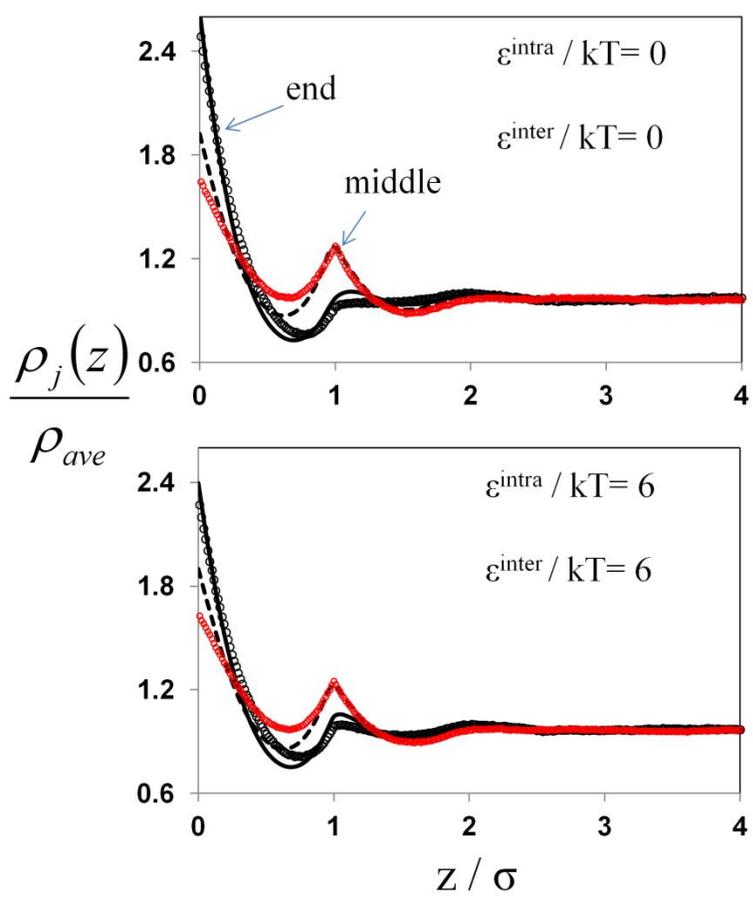

**Figure 7:**

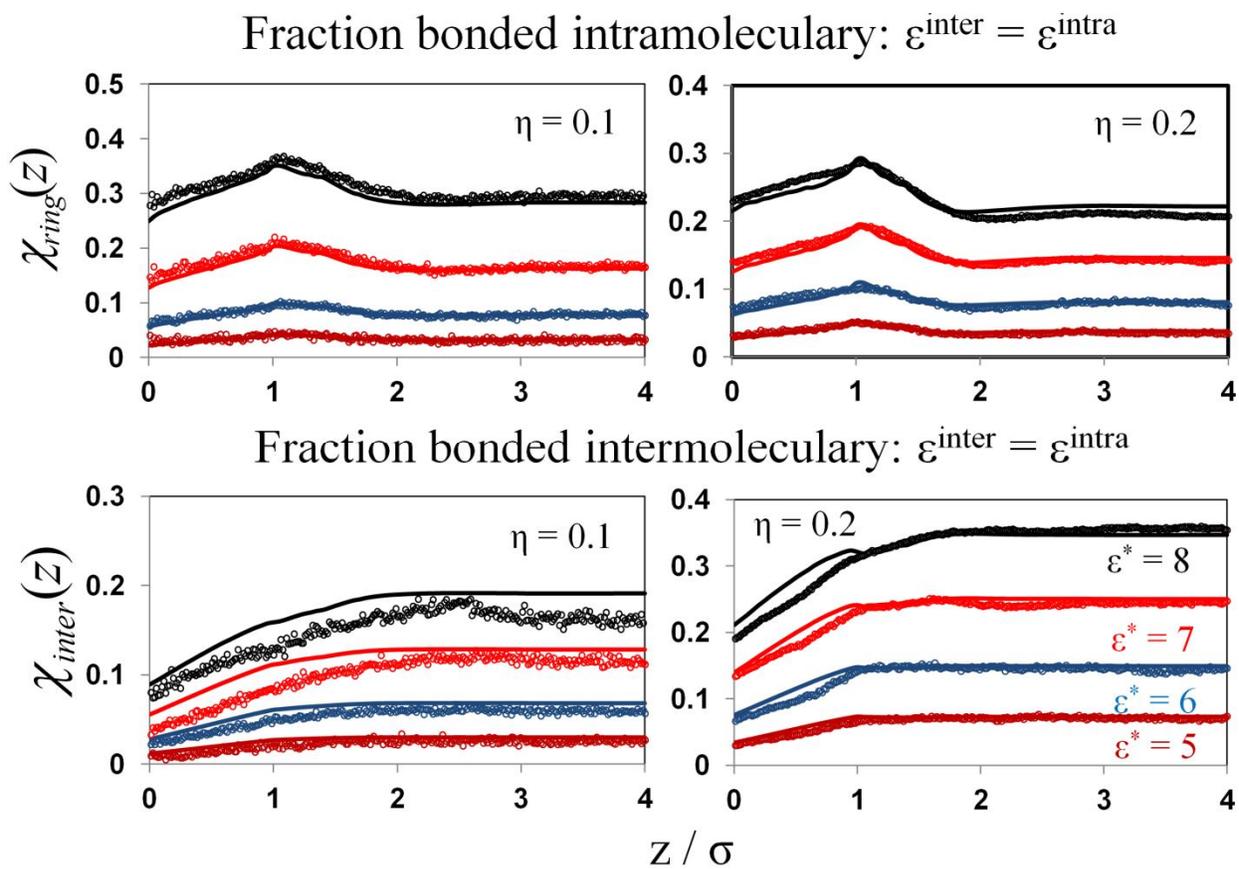

**Figure 8:**

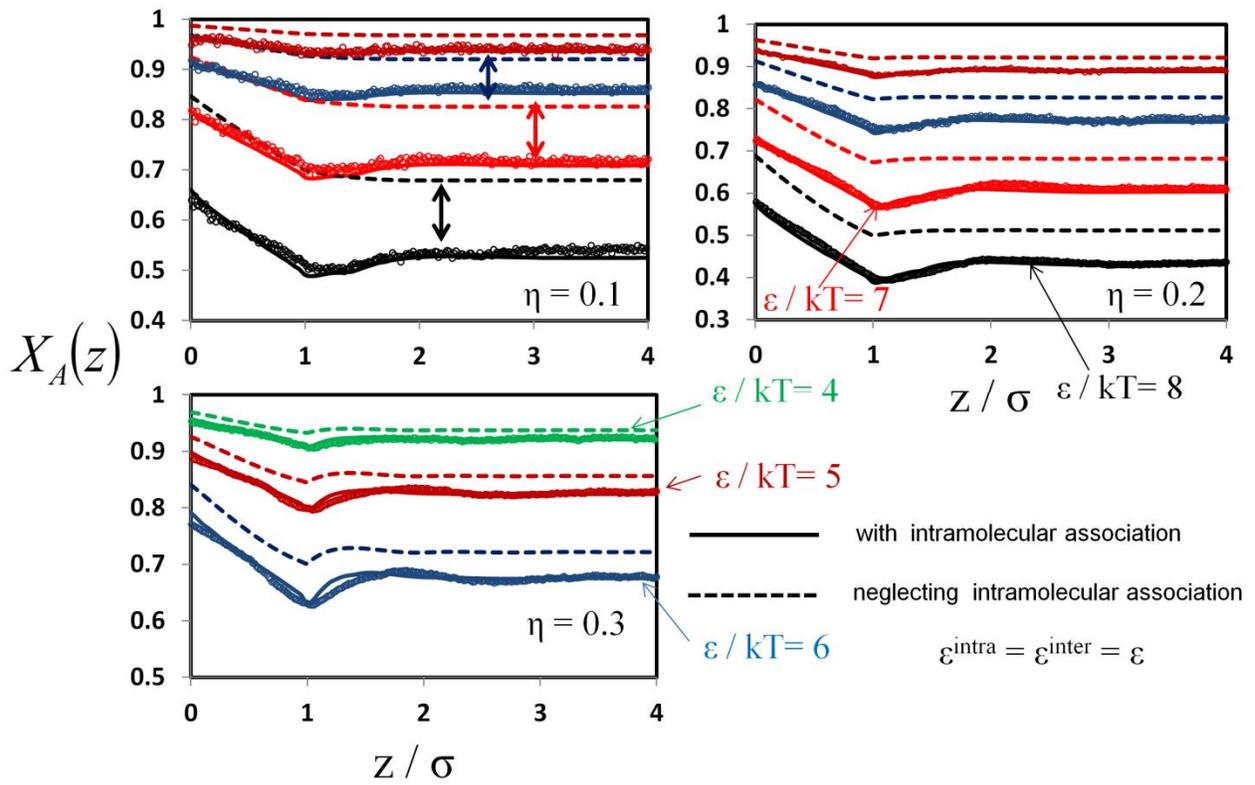

**Figure 9:**

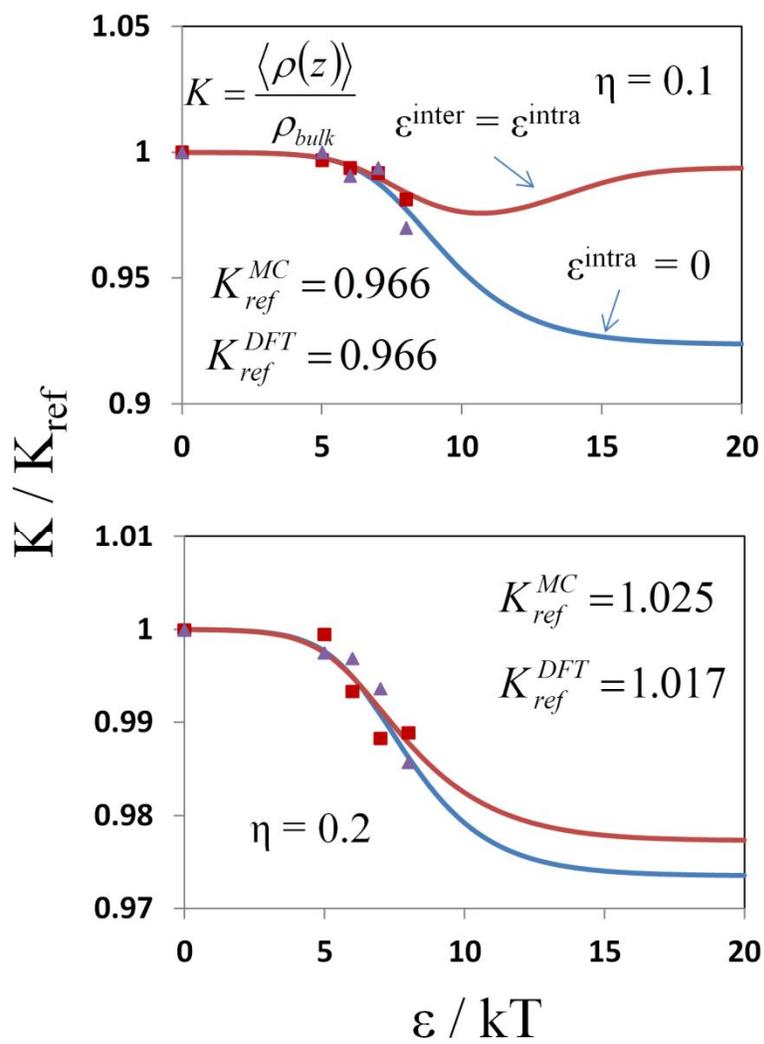

**Figure 10:**

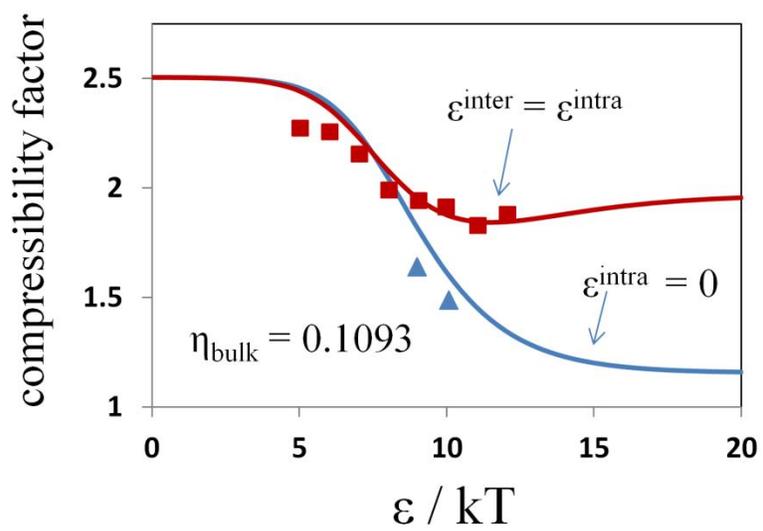

**Figure 11:**

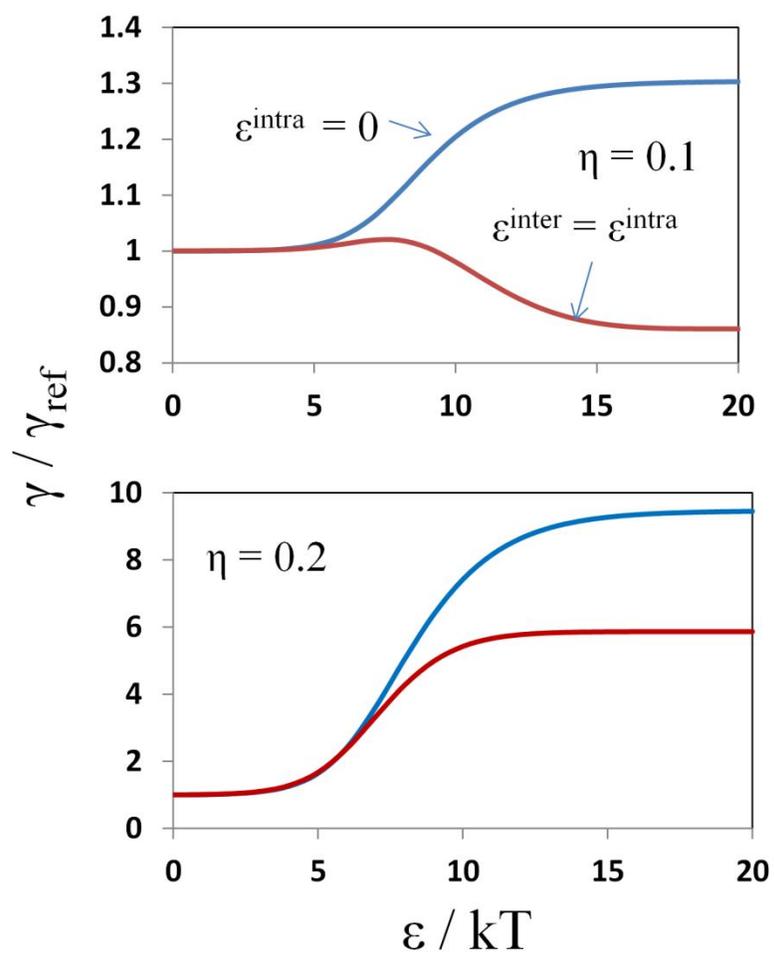

**Figure 12:**

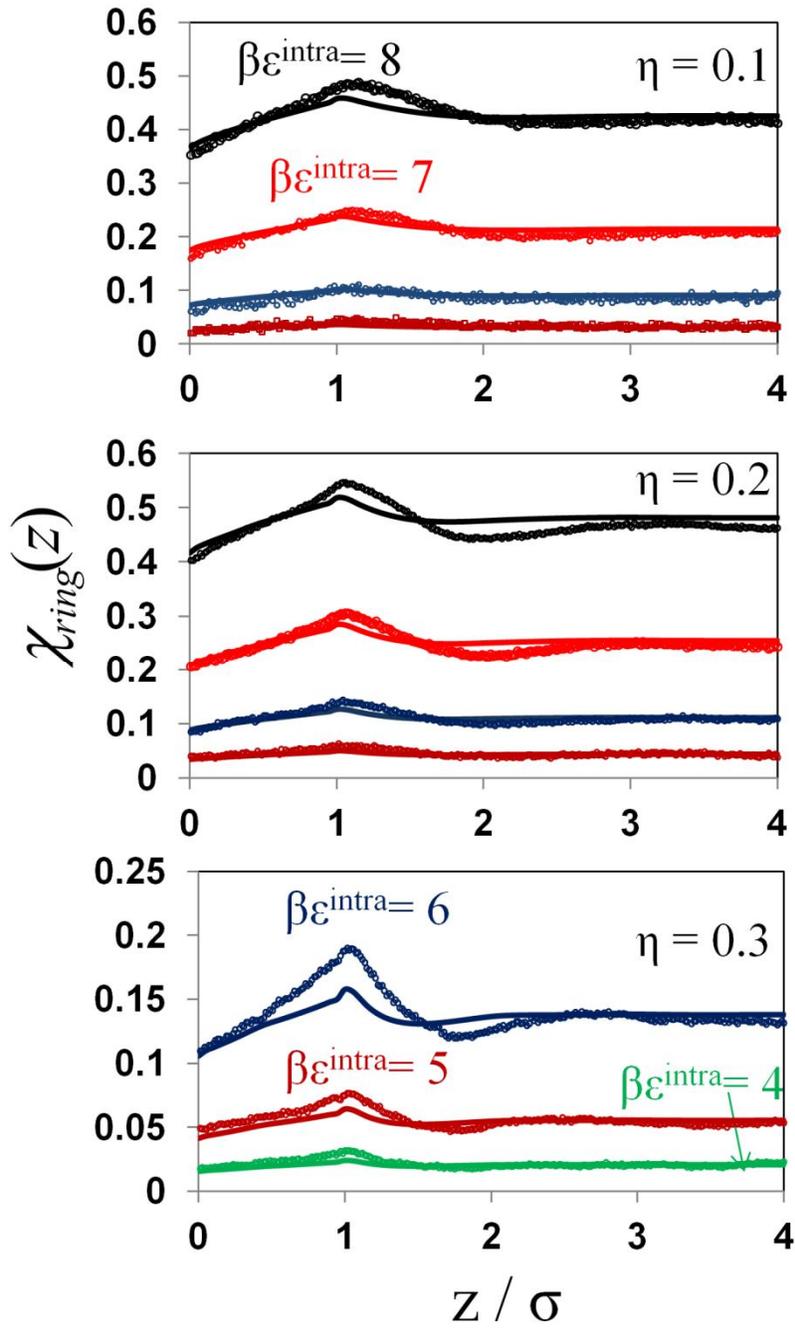

**Figure 13:**

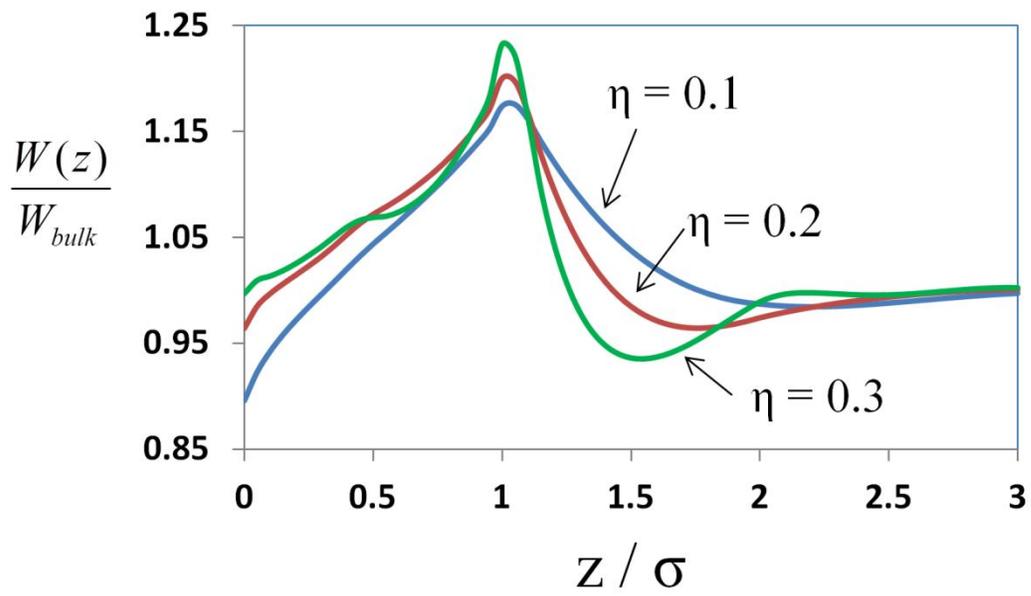

**Figure 14:**

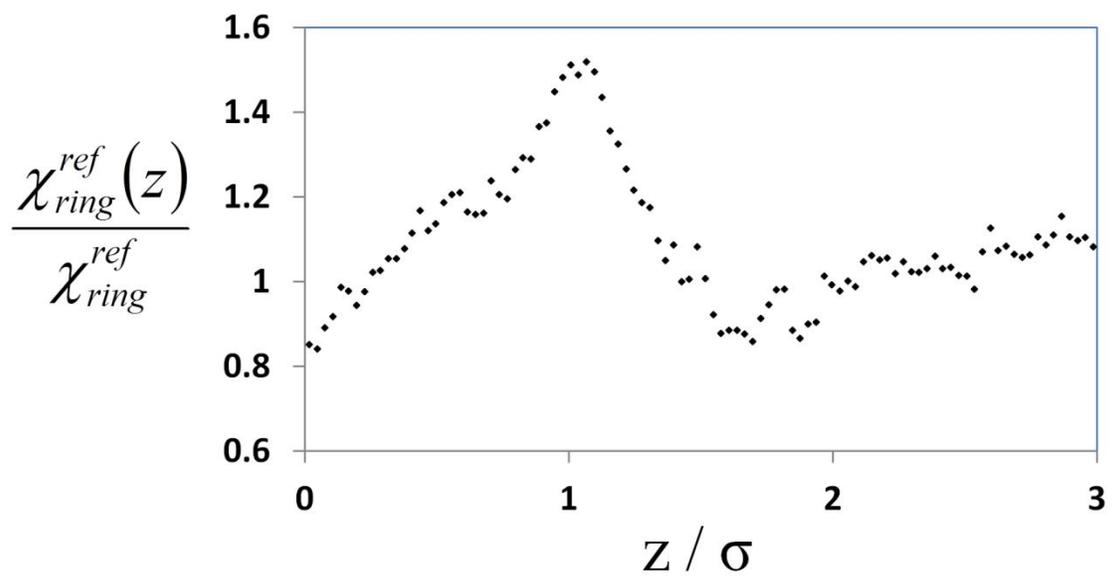

**Figure 15:**

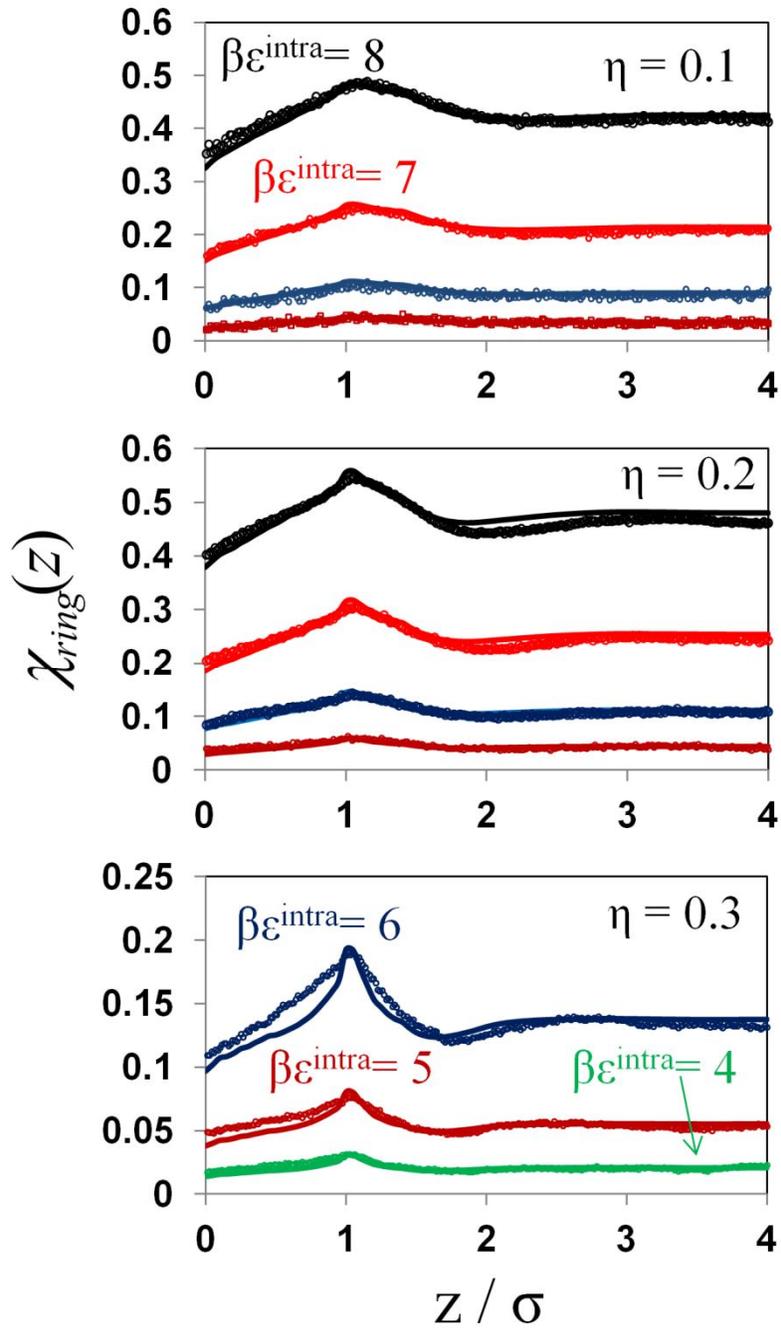

**Figure 16:**

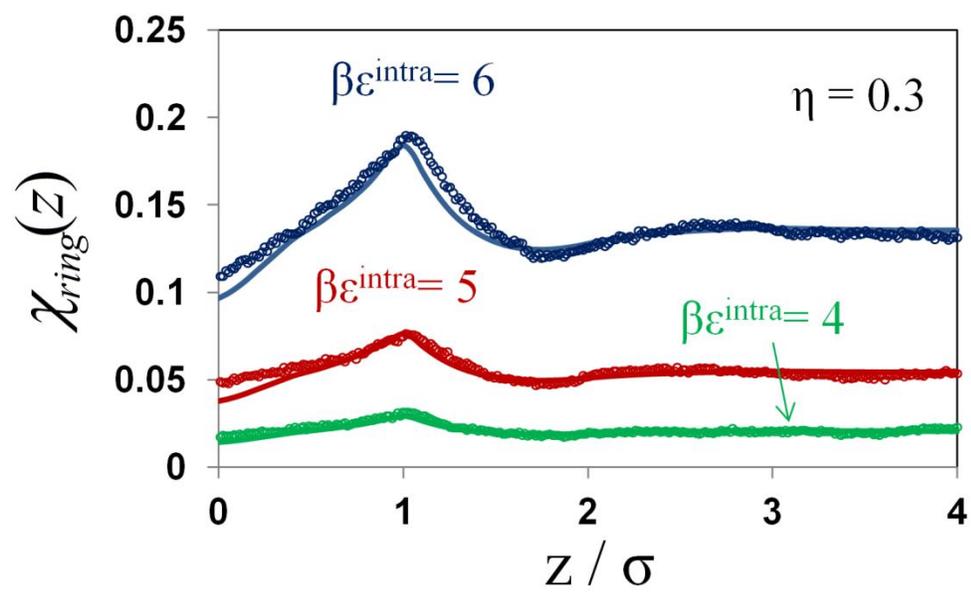